# Quantum Manipulation of Valleys in Bilayer Graphene: Theory and Applications

G. Y. Wu,[1,2,*] N.-Y. Lue,[2] and Y.-C. Chen[1]

Quantum manipulation of valleys in bilayer graphene is investigated. We establish an effective Schrodinger model, and identify two key mechanisms for valley manipulation – band structure warping and generalized valley-orbit interaction. Specifically, we implement valley qubits / FETs in bilayer graphene, as prospective quantum devices to build valley-based quantum / classical information processing.

[1]Department of Physics, National Tsing-Hua University, Hsin-Chu, Taiwan 30013, ROC
[2]Department of Electrical Engineering, National Tsing-Hua University, Hsin-Chu, Taiwan 30013, ROC
*Corresponding author. Email: yswu@ee.nthu.edu.tw

Graphene is a 2D material with immense potential for future electronics.[1-4] In particular, the binary-valued, valley degree of freedom (DOF) carried by a graphene electron opens up a new realm for electronics – valleytronics in graphene.[5-14] Valley filtering (or polarization)[5-8,10-12], Hall effects[6], optoelectronics[9], devices[5,8,13,14], and magnetic effects[6,13,15] have been investigated or proposed. Among interesting examples are qubits[13] / FETs[14], with the prospect to build valley-based information processing.

Valleytronic applications often require the presence of an energy gap to achieve valley-based device manipulation and/or attain quantum confinement of electron valleys. Our work considers AB-stacked bilayer graphene (BLG)[2-4], a gapped system under DC bias, and aims to establish a theory for quantum manipulation of a single electron valley, as well as implement crucial quantum devices in BLG based on the manipulation.[16] The utilization of an experimentally accessible system[3,17] here provides a major step towards the realization of these devices. An outline of the work is given below:
**a)** An effective Schrodinger model is presented as the theoretical framework.
**b)** Two key mechanisms – band warping and generalized valley-orbit interaction (gVOI) – are proposed for electrical valley manipulation.
**c)** Valley qubits in BLG for quantum computing / communications are demonstrated.
**d)** Valley FETs in BLG are illustrated, with potential advantages to build low-power, high-density ICs.

BLG shares with monolayer graphene the crucial property for valleytronics, e.g., the presence of a valley DOF, derived from a band structure with two degenerate and independent conduction band valleys (K and K') that transform into each other under time reversal symmetry. However, major differences exist between the two systems in the aspect of valley-dependent physics. In the monolayer case, the physics is underlain by the intralayer C-C hopping, which gives rise to the valley-orbit interaction (VOI) with an important role in valley manipulation.[13,14] In contrast, the physics in BLG is dominated by the interplay between the intra- and inter- layer couplings. As such, the VOI here takes a generalized form substantially different from that in monolayer graphene. In addition, valley-dependent trigonal warping appears in the BLG band structure, which not only enriches the valley-dependent physics but also provides, besides the VOI, a valuable mechanism for valley manipulation.

**(The Schrodinger Model)** The theory is based on the one-band Schrodinger model,

explained below. We denote the four C atoms in a unit cell as $A_1$, $B_1$ (in the 1st layer) and $A_2$, $B_2$ (in the 2nd layer), with $B_2$ right on top of $A_1$. A tight-binding model of BLG is characterized by the parameters $\Delta$, t, $\gamma_1$, and $\gamma_3$, with $2\Delta$ = *DC bias between the layers*, and t, $\gamma_1$, and $\gamma_3$ being, respectively, the *hopping between $A_1$ and $B_1$ (or $A_2$ and $B_2$), $A_1$ and $B_2$, and $B_1$ and $A_2$*. The band structure is summarized below.[2-4] An energy gap $2\Delta$ is opened in the band structure. Away from the gap, two distant bands are located at $\pm(\Delta^2+\gamma_1^2)^{1/2}$. In the limit where $\Delta \ll \gamma_1$, as assumed throughout this work, the full tight-binding description can be reduced, by the Schrieffer-Wolff type transformation, to the two-band model ($\hbar = 1$)[2]

$$\sum_{j=1,2} H_{ij}(k_x, k_y)\varphi_j = E\varphi_i, \quad i = 1, 2.$$

Here, $H_{ij}$ are Hamiltonian matrix elements and $(k_x, k_y)$ = *electron wave vector* relative to K / K' point. This equation describes the bands right near the gap, with the effect of distant states included through the perturbation theory. In the presence of a weak, slowly-varying external electric potential (V) with $\|V/\gamma_1\| \ll 1$, the effective mass theorem leads to the substitutions: $H_{ii} \to H_{ii} + V$, $(k_x, k_y) \to (-i\partial_x, -i\partial_y)$. As the above equation is a close analogy of the Dirac equation in gapped monolayer graphene, with $2\Delta$ being the corresponding "mass gap", we can further reduce it to a Schrodinger type equation, for $|(E\pm\Delta)/\Delta| \ll 1$ - the "nonrelativistic limit". For low-lying states of electrons, it gives

$$(E - \Delta)\varphi \approx [H_0(\tau) + H_1(\tau)]\varphi,$$

$$H_0(\tau) = -\frac{2\Delta}{\gamma_1^2}v_F^2 k^2 + \frac{1}{2\Delta}\left(\frac{1}{\gamma_1^2}v_F^4 k^4 + v'^2 k^2\right) + H_0^{(warping)}(\tau) + V,$$

$$H_1(\tau) = H_1^{(warping)}(\tau) + H_1^{(GVOI)}(\tau) + H_1',$$

$$H_0^{(warping)}(\tau) = \tau\frac{v'v_F^2}{\Delta\gamma_1}\left(3k_x^2 k_y - k_y^3\right),$$

$$H_1^{(warping)}(\tau) = \tau\frac{v'v_F^2}{\Delta\gamma_1}\left[\frac{2v_F^2}{\gamma_1^2}k^2 - \frac{1}{2\Delta^2}(\frac{1}{\gamma_1^2}v_F^4 k^4 + v'^2 k^2)\right](3k_x^2 k_y - k_y^3),$$

$$H_1^{(GVOI)} = \frac{1}{4\Delta^2}\Omega_\tau\{H_{21}[V, H_{12}]_-\}, \quad H_{12} \equiv \frac{1}{\gamma_1}v_F^2 k_+^2 + iv'k_-. \quad (1)$$

Here, we have taken the various "relativistic effect" ratios e.g., $\|V/\Delta\|$ and $|(E-\Delta)/\Delta|$, to be much less than unity, and retained only the leading- and next leading- order terms in the equation. The armchair direction is aligned along the x-axis, $v_F$ (Fermi velocity) = $3ta/2$ ("a" = *intralayer C-C distance*), and v' = $3\gamma_3 a/2$, $k_\pm \equiv k_x \pm i\tau k_y$, $\tau$ = *valley index* ("+ / -" for K / K'). We take t = 2.8eV, $\gamma_1$ = 0.4eV, and $\gamma_3$ = 0.3eV. $\Omega_\tau$

denotes operation acting upon the expression following it, by retaining only the τ-dependent terms in the expression.

Eqn. (1) gives the "Schrodinger Hamiltonian" $H_0$ and the 1$^{st}$-order "relativistic effect" $H_1$. $H_1'$ (not explicitly given) is the 1$^{st}$-order relativistic effect which is τ-independent and irrelevant to the present work. $H_0^{(warping)}$ and $H_1^{(warping)}$ above produce band warping. $H_1^{(gVOI)}$ is the gVOI, which reduces, for $v_F = 0$, to $\tau \frac{v'^2}{4\Delta^2} \nabla V \times \hat{k}$, the simple form of VOI[13,14,18] in monolayer graphene. Note that $H_0^{(warping)}$ and $H_1^{(gVOI)}$ are both τ-dependent and constitute key mechanisms for valley manipulation. Being valley-conserving, they are well suited to valley coherence-sensitive applications.

**(Two-Valley Qubits)** Fig. 1 shows a valley-based, two-electron qubit, which consists of a pair of laterally coupled quantum dots (QDs) of comparable sizes in BLG, in the (1, 1) charge configuration. The logical 0 / 1 states are represented by the two-valley singlet(S) / triplet($T_0$) states

$$|S> = \frac{1}{\sqrt{2}}(|K_L K'_R> - |K'_L K_R>), |T_0> = \frac{1}{\sqrt{2}}(|K_L K'_R> + |K'_L K_R>),$$

$$|K_L K'_R> \equiv c^+_{KL} c^+_{K'R} |vacuum>, |K'_L K_R> \equiv c^+_{K'L} c^+_{KR} |vacuum>.$$

$c^+$ is the electron creation operator, and $K_L$($K'_L$) and $K_R$($K'_R$) denote the ground states of K(K') valleys in the left and right QDs, respectively. The electron spins here are taken to be frozen in a triplet state by initialization, and can be dropped from the discussion.[19] The logical state space of the qubit is isomorphic to that of a spin-1/2 system, with the following correspondence

$$|S> \Leftrightarrow |\downarrow>, |T_0> \Leftrightarrow |\uparrow>, |K_L K'_R> \Leftrightarrow |\rightarrow>, |K'_L K_R> \Leftrightarrow |\leftarrow>,$$

where "↑", "↓", "→", and "←" denote "up", "down", "left", and "right" spin states quantized along the z- and x- axes of the spin system, respectively. The qubit/spin isomorphism provides a convenient framework to envision the qubit state transformation in terms of an effective spin rotation.

**(Warping-Based Two-Valley Manipulation)** An arbitrary qubit (or effective spin) rotation can be decomposed into two independent rotations, such as those around the x- and z- axes. We denote them as $R_x(\theta_x)$ and $R_z(\theta_z)$, respectively, where $\theta_x$ and $\theta_z$ are the angles of rotation. The corresponding qubit state transformations are given by

$$|K_L K'_R> \rightarrow e^{i\theta_x/2} |K_L K'_R>, \quad |K'_L K_R> \rightarrow e^{-i\theta_x/2} |K'_L K_R> \qquad (2)$$

for $R_x(\theta_x)$, and

$$|T_0> \to e^{i\theta_z/2}|T_0>, \quad |S> \to e^{-i\theta_z/2}|S> \tag{3}$$

for $R_z(\theta_z)$.

First, we discuss the generation of $R_x(\theta_x)$. As Eqn. (2) suggests, this may be enabled by introducing a valley-contrasting state evolution in the left QD (or both of the QDs),

$$|K_L> \to e^{i\theta_x/2}|K_L>, \quad |K'_L> \to e^{-i\theta_x/2}|K'_L>. \tag{2'}$$

The valley-contrasting phase $\theta_x$ in Eqn. (2') generally vanishes because $|K_L>$ and $|K'_L>$ usually evolve with the same phase due to the valley degeneracy. For a finite $\theta_x$, the symmetry between $|K_L>$ and $|K'_L>$ must be broken. The condition of breaking is discussed below.

**i) warping-based valley symmetry breaking**

Valley symmetry breaking can be achieved by applying to the QD an AC electric field, $\varepsilon_{ac}\sin w_s t$, in the y-direction, as follows. We ignore the relativistic effect, and write the wave equation for the QD state ($\psi$)

$$H_0(t;\tau)\psi(x,y,t;\tau) = i\partial_t\psi(x,y,t;\tau), \tag{4}$$
$$V = V_{QD} + e\varepsilon_{ac}y\sin w_s t.$$

Here, $V_{QD}$ is the QD confinement potential, and the time dependence of $H_0$ derives from the AC field. The introduction of AC field breaks time reversal symmetry. However, the presence of $H_0^{(warping)}$ in $H_0$ is indispensable, for the condition of breaking. Without the term, $H_0$ would be $\tau$-independent, and the states of $\tau = \pm$, e.g., $|K_L>$ and $|K'_L>$, would evolve in a $\tau$-independent fashion, giving a vanishing $\theta_x$.

**ii) valley-contrasting geometric phase**

We apply Eqn. (4) to the QD ground state. For $\varepsilon_{ac} = 0$, the ground state solution of Eqn. (4) is denoted as $\psi_0(x, y, t; \tau)$, with $w_0$ (energy relative to the conduction band edge). Note that $\psi_0(x, y, t; +) = |K_L>$, $\psi_0(x, y, t; -) = |K'_L>$, and $w_0$ is $\tau$-independent due to time reversal symmetry. For $\varepsilon_{ac} \neq 0$, we consider the simple case where $V_{QD} = \frac{1}{2} m w_0^2 (x^2 + y^2)$ ("m" = a mass parameter) and the AC field is weak and quasi-static with $w_s \ll w_0$. To the leading order in $\varepsilon_{ac}$, we write the total potential $V \approx V_{QD}(x, y + y_0(t))$, which describes a dynamical QD with "$-y_0(t) \equiv -e\varepsilon_{ac}\sin w_s t / m w_0^2$" being the dynamical equilibrium position. This admits the following simple adiabatic solution[20]

$$\psi(x,y,t;\tau) \approx \psi_0(x, y+y_0(t), t;\tau)\exp\left[-i\int^t \gamma_0(\tau)dt\right], \tag{5}$$
$$\text{(geometric phase)}$$

$$\psi_0(x, y+y_0(t), t; \tau) = \varphi_0(x, y+y_0(t); \tau) \exp\left(-i\int^t w_0 dt\right),$$
$$\text{(dynamical phase)}$$

$$\gamma_0(\tau) \approx (\partial_t y_0(t)) <\varphi_0(x,y;\tau)| k_y \varphi_0(x,y;\tau)>.$$

An intuitive interpretation follows. It is the instantaneous ground state ("$\psi_0$", with spatial part = "$\varphi_0$", and temporal part = "dynamical phase") observed at the moving QD and transformed back to the laboratory reference frame. The transformation produces an energy shift $<\partial_t y_0 k_y>$ (denoted $\gamma_0$) which generates the geometric phase. This phase (or $\gamma_0$) is $\tau$-dependent, as shall be shown in **iii)**, and therefore is identified with $\theta_x/2$ in Eqns. (2') or (2).

**iii) qubit rotation rate**

With $\gamma_0$ generating the geometric phase, the typical qubit rotation rate is given by $\gamma_0$. We derive $\gamma_0$ with the perturbation theory, treating the warping term in $H_0$ as a perturbation. To the leading order, this yields

$$\gamma_0(\tau) \approx (2\partial_t y_0(t)) \sum_n \frac{<0| H_0^{(warping)}(\tau)|2n+1><2n+1|k_y|0>}{w_0^{(0)} - w_{2n+1}^{(0)}},$$

$$H_0^{(0)} |n> = w_n^{(0)} |n>,$$

$$H_0^{(0)} \equiv -\frac{2\Delta}{\gamma_1^2} v_f^2 k^2 + \frac{1}{2\Delta}\left(\frac{1}{\gamma_1^2} v_F^4 k^4 + v'^2 k^2\right) + V_{QD}. \tag{6}$$

$H_0^{(0)}$ is the QD Hamiltonian ($H_0$) with the warping term removed. The above expression shows an explicit $\tau$-dependence in $\gamma_0$ derived from the warping term. For an order-of-magnitude estimate, we take L (QD size) ~ 300A, $\Delta$ ~ 5 meV, $\pi/w_s$ ~ 0.1ns, $\hbar w_0$ ~ $m w_0^2 L^2$ ~ O(meV), and $e\varepsilon_{ac} / m w_0^2$ ~ 0.3 L. This yields $\gamma_0$ ~ GHz, leading to the prospect of operating valley qubits in the GHz range.

The two-valley qubit is analogous to a two-spin qubit[21] and, hence, shares advantages of the latter in several aspects, summarized below.[13] First, it has a decoherence-free logical state space. Second, $R_z$ can be generated by utilizing the exchange coupling (denoted "J") between the two localized electrons in the qubit. With $J \sim 4t^2 / U$ (t = *interdot tunneling* and U = *on-site Coulomb repulsion*), this gives a way to control $R_z$, e.g., by tuning the interdot tunneling. Last, two valley qubits can be placed side by side to form a CPHASE gate. $R_x$, $R_z$ and the CPHASE gate constitute universal quantum computing.[22-24]

We note an alternative method to generate $R_x$. It places the QDs in a normal magnetic field to induce valley asymmetry and applies an in-plane DC electric field to the QDs for qubit rotation. This method also involves the warping effect, and will be reported elsewhere.

**(Valley FETs)** Fig. 2(a) shows a valley FET in gapped BLG. The source and drain are AGNRs (armchair graphene nanoribbons), and the channel is a graphene QW (quantum wire) aligned in the armchair direction and subject to the control of a side gate bias. The bias produces in the channel an in-plane electric field in the y-direction.

In both the leads, the unique boundary condition in an AGNR mixes K and K' valleys in a 50-50 ratio, giving the lead subband state as symbolically represented by $(|K\rangle + S|K'\rangle)/2^{1/2}$[25], where $S = \pm 1$ is subband dependent. This determines the specific valley polarization injected into the channel and detected at the drain.

**(GVOI-Based Valley Precession)** The gVOI in the channel provides a key mechanism to switch on / off the FET with the gate bias, as follows. Under the bias, the energy subbands in the channel are valley-split due to the gVOI (as shown below), giving a wave vector difference, "$k_+ - k_-$", between the states of $\tau = \pm$, as shown in Fig. 2(b). Therefore, after a source electron is injected into the channel, the two valley components in the electron evolve with different phases, leading to the channel state $(e^{i\varphi(x)}|K\rangle + S|K'\rangle)/2^{1/2}$. The phase difference $\varphi(x) = (k_+ - k_-)x$ here increases linearly with the distance "x" travelled. This describes an electron precession in the valley space, with $\varphi$ = precession angle. At the end of the channel (x = L), depending on $\varphi = 2n\pi$ [$(2n+1)\pi$], the channel and drain polarizations are parallel [orthogonal], admitting [blocking] the electron into[off] the drain. The gVOI here plays a role similar to the spin-orbit interaction (SOI) utilized in a spin FET[26], to achieve the on-off switch function.

We derive the gVOI-based valley splitting / precession in the channel. In the present case, we write $H_1^{(gVOI)}$ in Eqn. (1) explicitly, with

$$H_1^{(GVOI)} = i\tau \frac{v'^2}{4\Delta^2} k_x \underline{k_y V} - i\tau \frac{v_F^4}{2\Delta^2 \gamma_1^2} k_x \left( \underline{k_y^2 V k_y} + \underline{k_y V k^2} \right) + \tau \frac{v' v_F^2}{4\Delta^2 \gamma_1} \left[ \underline{k_y^3 V} + 2\underline{k_y^2 V k_y} \right]$$

The underlined expressions in $H_1^{(gVOI)}$ are evaluated first. Here, we take $k_x$ = *electron wave vector* along the channel, and the potential V to be y-dependent only, e.g., $V = V_{QW}(y) + e\varepsilon_y y$, with $V_{QW}(y) = \frac{1}{2} mw_0^2 y^2 - Dy^4$ being the QW confinement potential. $\varepsilon_y$ = *gate field* applied, $w_0$ = *subband edge*, and the quartic term, "$-Dy^4$", is introduced such that $V_{QW}$ simulates a realistic, finite confinement potential which flattens out at distant y. Without the quartic term, the gate field would only shift $V_{QW}$ to a new equilibrium position without producing any effect.

The valley degeneracy is lifted by $\varepsilon_y$ due to the gVOI. This can be deduced by a symmetry argument based on $H_1^{(gVOI)}$ - The odd-in-$k_x$, $\tau$-dependent terms in $H_1^{(gVOI)}$ remove the valley degeneracy at a given $k_x$. For a quantitative estimate of the splitting, we take $H_0^{(0)}$ and $|n\rangle$ in the qubit discussion (with the replacement $V_{QD} \to \frac{1}{2} mw_0^2 y^2$ for the QW confinement) as the unperturbed Hamiltonian and eigenstates, respectively, and "$e\varepsilon_y y$", "$Dy^4$", and $H_1^{(gVOI)}$ as perturbations.[27] This yields ($m^* \equiv$ *subband effective mass*)

$$k_+ - k_- = -\frac{2m^* \alpha_{vo}}{\hbar^2}, \quad \varphi = -\frac{2m^* \alpha_{vo}}{\hbar^2} L,$$

$$\alpha_{vo}/\varepsilon_y \approx \frac{-ev_F^4 D}{\Delta^2 \gamma_1^2} \sum_n' \langle 0|k_y^2|2n\rangle \langle 2n|y^4|0\rangle / \left(w_0^{(0)} - w_{2n}^{(0)}\right)$$

$$+ \frac{2eD}{\Delta^2} \sum_n \langle 0| \left[ \frac{2v_F^4}{\gamma_1^2} \left(3y^2 k_y + y^3 k_y^2\right) + v'^2 y^3 \right] |2n+1\rangle \langle 2n+1|y|0\rangle / \left(w_0^{(0)} - w_{2n+1}^{(0)}\right), \quad (7)$$

where $\alpha_{vo}$ is the "Rashba constant" due to the gVOI. As shown above, $\alpha_{vo}$ is $\varepsilon_y$-dependent, and this permits the electric control of valley precession angle. We estimate $\alpha_{vo}$ using the following parameters: W (QW width) ~ 300Å, $\Delta$ ~ 5 meV, and $\hbar w_0$ ~ $m w_0^2 W^2$ ~ $e\varepsilon_y W$ ~ $DW^4$ ~ O(meV). This yields $\alpha_{vo}$ in the range $10^{-12} - 10^{-11}$ eV-m, comparable to the large, SOI-caused Rashba constant in InAs[28]. By analogy to spin FETs, valley FETs thus carry similar potential advantages in building low-power, high-density ICs.

Finally, we briefly note the issue of valley coherence in BLG. In the case of monolayer graphene, estimation[13,14] shows a reasonable coherence time for valley manipulation. As the decoherence mechanism in both cases is dominated by the intervalley[29] K ↔ K' scattering, which is in-plane in nature, we expect a similar estimate for BLG, due to analogous in-plane properties in the two cases.[30]

**Acknowledgment** − We would like to thank the support of ROC National Science Council through the Contract No. NSC101-2112-M-007-012.

# Figure Captions

1. The valley pair qubit consisting of laterally coupled QDs of comparable sizes in BLG. $V_\Delta$, $V_L$, $V_R$, and $V_C$ are electrical gates, which may be arranged in a mirror-symmetric fashion as suggested here (only those above the graphene layers are explicitly shown). DC biases applied to these gates open energy gaps in the bilayer system and define the QDs. AC biases applied to $V_L$ / $V_R$ generate $R_x$. $V_C$ controls the interdot tunneling "t" (or the exchange coupling "J"), and hence $R_z$.

2. **(a)** A schematic plot of the valley FET in BLG, with AGNR source and drain, and a graphene QW channel (aligned in the armchair direction) subject to the control of a side gate bias. In order to define the QW, the side gates and / or vertical gates (such as $V_\Delta$ in Fig. 1 but not shown here) can be arranged in a mirror-symmetric fashion, as in Fig. 1, with DC biases applied upon them. Electron states in the various regions are also shown. **(b)** Under the gate bias, energy subbands in the channel are valley-split due to the gVOI, giving a wave vector difference, "$k_+ - k_-$", between the states of $\tau = \pm$.

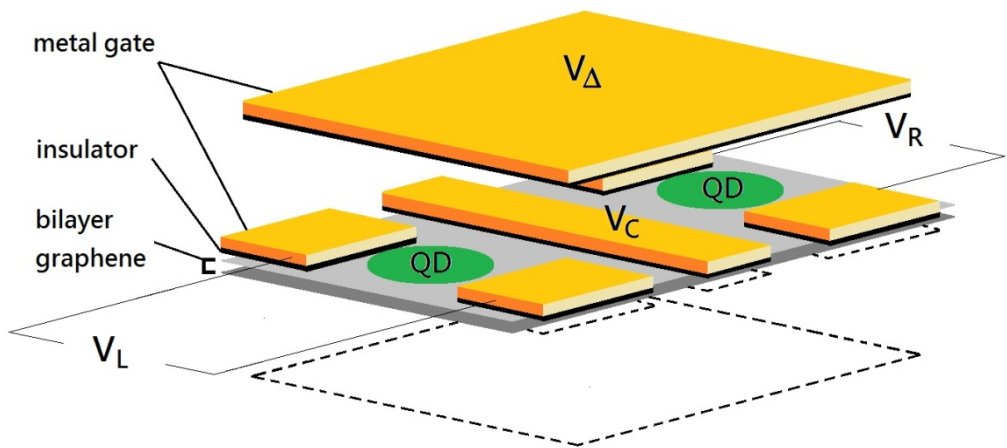

Fig. 1

(a)

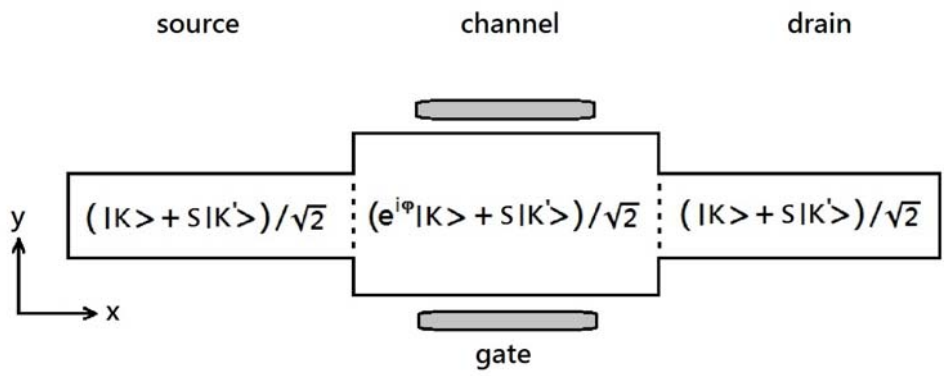

(b)

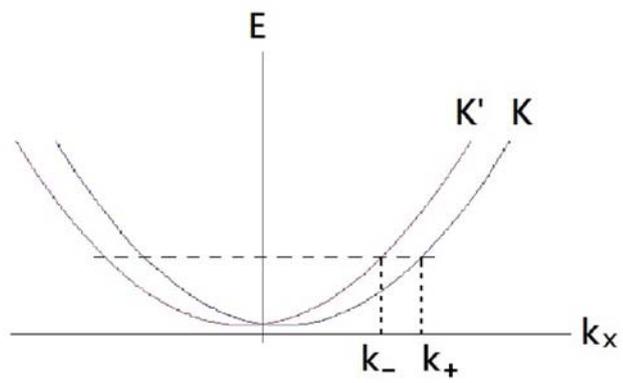

Fig. 2